\documentclass[aoas,nameyear,seceqn,dvips]{arximspdf}
\usepackage{dcolumn}
\usepackage{graphicx}

\doi{10.1214/09-AOAS254}
\volume{3}
\issue{4}
\pubyear{2009}
\firstpage{1738}
\lastpage{1757}

\makeatletter
\newtheorem{Theo}{Theorem}
\newcommand{\widehatt}{\hat}
\newcommand{\wPE}{\widehat{\operatorname{PE}}}
\newcolumntype{d}[1]{D{.}{.}{#1}}
\DeclareMathAlphabet\mathcaligr{OMS}{cmsy}{m}{n}
\makeatother

\begin{document}
\begin{frontmatter}

\title{Structured variable selection and estimation}
\runtitle{Structured variable selection and estimation}

\begin{aug}
\author[A]{\fnms{Ming} \snm{Yuan}\thanksref{t1}\ead[label=e1]{myuan@isye.gatech.edu}\corref{}},
\author[A]{\fnms{V. Roshan} \snm{Joseph}\thanksref{t2}\ead[label=e2]{roshan@isye.gatech.edu}}
\and
\author[B]{\fnms{Hui} \snm{Zou}\thanksref{t3}\ead[label=e3]{hzou@stat.umn.edu}}
\runauthor{M. Yuan, V. R. Joseph and H. Zou}
\affiliation{Georgia Institute of Technology, Georgia Institute of
Technology and~University~of~Minnesota}
\address[A]{
M. Yuan\\
V. R. Joseph\\
School of Industrial and Systems Engineering\\
Georgia Institute of Technology\\
755 Ferst Drive NW\\
Atlanta, Georgia 30332\\
USA\\
\printead{e1}\\
\phantom{E-mail:\ }\printead*{e2}
} 
\address[B]{
H. Zou\\
School of Statistics\\
University of Minnesota\\
224 Church St. SE\\
Minneapolis, Minnesota 55455\\
USA\\
\printead{e3}
}

\thankstext{t1}{Supported in part by NSF Grants DMS-MPSA-06-24841 and
DMS-07-06724.}
\thankstext{t2}{Supported in part by NSF Grant CMMI-0448774.}
\thankstext{t3}{Supported in part by NSF Grant DMS-07-06733.}
\end{aug}

\received{\smonth{10} \syear{2008}}
\revised{\smonth{5} \syear{2009}}

%
\begin{abstract}
In linear regression problems with related predictors, it is desirable
to do variable selection and estimation by maintaining the hierarchical
or structural relationships among predictors. In this paper we propose
non-negative garrote methods that can naturally incorporate such
relationships defined through effect heredity principles or marginality
principles. We show that the methods are very easy to compute and enjoy
nice theoretical properties. We also show that the methods can be
easily extended to deal with more general regression problems such as
generalized linear models. Simulations and real examples are used to
illustrate the merits of the proposed methods.
\end{abstract}

%

\end{frontmatter}

\section{Introduction}

When considering regression with a large number of predictors, variable
selection becomes important. Numerous methods have been proposed in the
literature for the purpose of variable selection, ranging from the
classical information criteria such as AIC and BIC to regularization
based modern techniques such as the nonnegative garrote [Breiman
(\citeyear{Breiman1995})], the Lasso [Tibshirani (\citeyear{Tibshirani1996})]
and the SCAD [Fan and Li (\citeyear{Fan2001})],
among many others. Although these methods enjoy excellent performance
in many applications, they do not take the hierarchical or structural
relationship among predictors into account and therefore can lead to
models that are hard to interpret.

Consider, for example, multiple linear regression with both main
effects and two-way interactions where a dependent variable $Y$ and $q$
explanatory variables $X_1, X_2, \ldots, X_q$ are related through
%
\begin{equation}
\label{2way}
Y = \beta_1X_1+\cdots+\beta_qX_q+\beta_{11}X_1^2+\beta_{12}X_1X_2
+ \cdots+\beta_{qq}X_q^2+\varepsilon,
\end{equation}
where $\varepsilon\sim\mathcaligr N(0, \sigma^2)$. Commonly used general
purpose variable selection techniques, including those mentioned above,
do not distinguish interactions $X_iX_j$ from main effects $X_i$ and
can select a model with an interaction but neither of its main effects,
that is, $\beta_{ij}\neq0$ and $\beta_i=\beta_j=0$. It is therefore
useful to invoke the so-called effect heredity principle [Hamada and Wu
(\citeyear{Hamada1992})] in this situation. There are two popular versions of the
heredity principle [Chipman (\citeyear{Chipman1996})]. Under \textit{strong heredity}, for
a two-factor interaction effect $X_iX_j$ to be active both its parent
effects, $X_i$ and $X_j$, should be active; whereas under \textit{weak
heredity} only one of its parent effects needs to be active. Likewise,
one may also require that $X_i^2$ can be active only if $X_i$ is also
active. The strong heredity principle is closely related to the notion
of marginality [Nelder (\citeyear{Nelder1977}), McCullagh and Nelder (\citeyear{McCullagh1989}), Nelder
(\citeyear{Nelder1994})] which ensures that the response surface is invariant under
scaling and translation of the explanatory variables in the model.
Interested readers are also referred to McCullagh (\citeyear{McCullagh2002}) for a rigorous
discussion about what criteria a sensible statistical model should
obey. Li, Sudarsanam and Frey (\citeyear{Li2006}) recently conducted a meta-analysis
of 113 data sets from published factorial experiments and concluded
that an overwhelming majority of these real studies conform with the
heredity principles. This clearly shows the importance of using these
principles in practice.

These two heredity concepts can be extended to describe more general
hierarchical structure among predictors. With slight abuse of notation,
write a general multiple linear regression as
%
\begin{equation}
\label{1.1}
Y = X \beta+ \varepsilon,
\end{equation}
where $X=(X_1, X_2, \ldots, X_p)$ and $\beta= (\beta_1, \ldots,\beta
_p)^\prime$. Throughout this paper, we center each variable so that
the observed mean is zero and, therefore, the regression equation has
no intercept. In its most general form, the hierarchical relationship
among predictors can be represented by sets $\{\mathcaligr D_{i}\dvtx
i=1,\ldots
,p\}$, where $\mathcaligr D_i$ contains the parent effects of the $i$th
predictor. For example, the dependence set of $X_iX_j$ is $\{X_i,X_j\}$
in the quadratic model~(\ref{2way}). In order that the $i$th variable
can be considered for inclusion, all elements of $\mathcaligr D_i$ must be
included under the strong heredity principle, and at least one element
of $\mathcaligr D_i$ should be included under the weak heredity principle.
Other types of heredity principles, such as the partial heredity
principle [Nelder (\citeyear{Nelder1998})], can also be incorporated in this framework.
The readers are referred to Yuan, Joseph and Lin (\citeyear{Yuan2007}) for further
details. As pointed out by Turlach (\citeyear{Turlach2004}), it could be very challenging
to conform with the hierarchical structure in the popular variable
selection methods. In this paper we specifically address this issue and
consider how to effectively impose such hierarchical structures among
the predictors in variable selection and coefficient estimation, which
we refer to as \textit{structured variable selection and estimation}.

Despite its great practical importance, structured variable selection
and estimation has received only scant attention in the literature.
Earlier interests in structured variable selection come from the
analysis of designed experiments where heredity principles have proven
to be powerful tools in resolving complex aliasing patterns. Hamada and
Wu (\citeyear{Hamada1992}) introduced a modified stepwise variable selection procedure
that can enforce effect heredity principles. Later, Chipman (\citeyear{Chipman1996}) and
Chipman, Hamada and Wu (\citeyear{Chipman1997}) discussed how the effect heredity can be
accommodated in the stochastic search variable selection method
developed by George and McCulloch (\citeyear{George1993}). See also Joseph and Delaney
(\citeyear{Joseph2007}) for another Bayesian approach. Despite its elegance, the
Bayesian approach can be computationally demanding for large scale
problems. Recently, Yuan, Joseph and Lin (\citeyear{Yuan2007}) proposed generalized
LARS algorithms [Osborne, Presnell and Turlach (\citeyear{Osborne2000}), Efron et al.
(\citeyear{Efron2004})] to incorporate heredity principles into model selection. Efron
et al. (\citeyear{Efron2004}) and Turlach (\citeyear{Turlach2004}) also considered alternative strategies
to enforce the strong heredity principle in the LARS algorithm.
Compared with earlier proposals, the generalized LARS procedures enjoy
tremendous computational advantages, which make them particularly
suitable for problems of moderate or large dimensions. However, Yuan
and Lin (\citeyear{Yuan2007}) recently showed that LARS may not be consistent in
variable selection. Moreover, the generalized LARS approach is not
flexible enough to incorporate many of the hierarchical structures
among predictors. More recently, Zhao, Rocha and Yu (\citeyear{Zhao2006}) and Choi, Li
and Zhu (\citeyear{Choi2006}) proposed penalization methods to enforce the strong
heredity principle in fitting a linear regression model. However, it is
not clear how to generalize them to handle more general heredity
structures and their theoretical properties remain unknown.

In this paper we propose a new framework for structured variable
selection and estimation that complements and improves over the
existing approaches. We introduce a family of shrinkage estimator that
is similar in spirit to the nonnegative garrote, which Yuan and Lin
(\citeyear{YuanLin2007}) recently showed to enjoy great computational advantages, nice
asymptotic properties and excellent finite sample performance. We
propose to incorporate structural relationships among predictors as
linear inequality constraints on the corresponding shrinkage factors.
The resulting estimates can be obtained as the solution of a quadratic
program and very efficiently solved using the standard quadratic
programming techniques. We show that, unlike LARS, it is consistent in
both variable selection and estimation provided that the true model has
such structures. Moreover, the linear inequality constraints can be
easily modified to adapt to any situation arising in practical problems
and therefore is much more flexible than the existing approaches. We
also extend the original nonnegative garrote as well as the proposed
structured variable selection and estimation methods to deal with the
generalized linear models.

The proposed approach is much more flexible than the generalized LARS
approach in Yuan, Joseph and Lin (\citeyear{Yuan2007}). For example, suppose a group of
variables is expected to follow strong heredity and another group weak
heredity, then in the proposed approach we only need to use the
corresponding constraints for strong and weak heredity in solving the
quadratic program, whereas the approach of Yuan, Joseph and Lin (\citeyear{Yuan2007}) is
algorithmic and therefore requires a considerable amount of expertise
with the generalized LARS code to implement these special needs.
However, there is a price to be paid for this added flexibility: it is
not as fast as the generalized LARS.

The rest of the paper is organized as follows. We introduce the
methodology and study its asymptotic properties in the next section. In
Section~\ref{sec3} we extend the methodology to generalized linear models.
Section~\ref{sec4} discusses the computational issues involved in the
estimation. Simulations and real data examples are presented in
Sections~\ref{sec5} and \ref{sec6} to illustrate the proposed methods. We conclude with
some discussions in Section~\ref{sec7}.

\section{Structured variable selection and estimation}\label{sec2}
The original nonnegative garrote estimator introduced by Breiman (\citeyear{Breiman1995})
is a scaled version of the least square estimate. Given $n$ independent
copies $(\mathbf{x}_1,y_1),\ldots,(\mathbf{x}_n,y_n)$ of $(X,Y)$
where~$X$ is a
$p$-dimensional covariate and $Y$ is a response variable, the shrinkage
factor $\theta(M)=(\theta_1(M),\ldots, \theta_p(M))'$ is given as
the minimizer to
%
\begin{equation}
\label{shrink}
\frac{1}{2}\|Y-Z\theta\|^{2},\qquad \mbox{subject to } \sum
_{j=1}^p \theta_j\le M \mbox{ and } \theta_j\ge0\ \forall j,
\end{equation}
where, with slight abuse of notation, $Y=(y_1,\ldots,y_n)'$,
$Z=(\mathbf{z}
_1,\ldots,\mathbf{z}_n)'$, and~$\mathbf{z}_i$ is a $p$ dimensional
vector whose
$j$th element is $x_{ij}\widehatt{\beta}^\mathrm{LS}_j$ and\vspace*{1pt} $\widehatt
{\beta}^\mathrm{LS}$ is the least square estimate based on~(\ref{1.1}).
Here $M\ge0$ is a tuning parameter. The nonnegative garrote estimate
of the regression\vspace*{1pt} coefficient is subsequently defined as $\widehatt
{\beta}^\mathrm{NG}_j(M)=\theta_j(M)\widehatt{\beta}^\mathrm{LS}_j$,
$j=1,\ldots,p$. With an appropriately chosen tuning parameter $M$, some of
the scaling factors could be estimated by exact zero and, therefore,
the corresponding predictors are eliminated from the selected model.

\subsection{Strong heredity principles}

\begin{figure}

\includegraphics{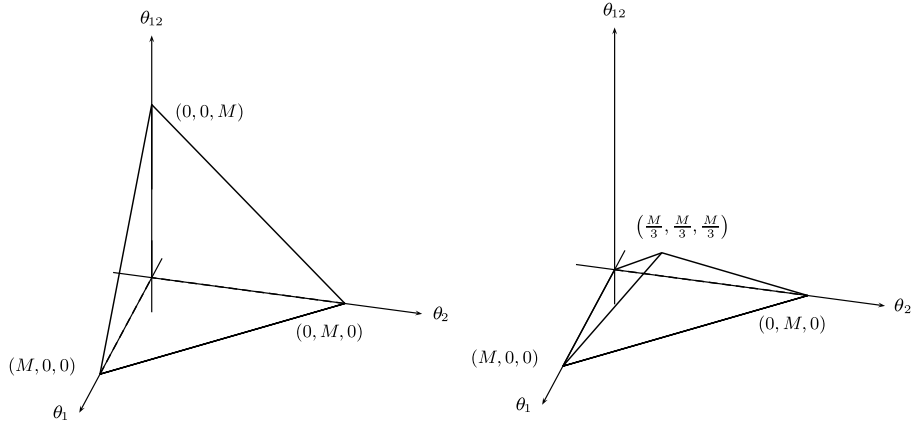}

\caption{Feasible region of the nonnegative garrote with (right) and
without (left) strong heredity constraints.}
\label{fig:strong}
\end{figure}

Following Yuan, Joseph and Lin (\citeyear{Yuan2007}), let~$\mathcaligr D_i$ contain the
parent effects of the $i$th predictor. Under the strong heredity
principle, we need to impose the constraint that $\widehatt{\beta
}_j\neq0$ for any $j\in\mathcaligr D_i$ if $\widehatt{\beta_i}\neq
0$. A
naive approach to incorporating effect heredity is therefore to
minimize~(\ref{shrink}) under this additional constraint. However, in
doing so, we lose the convexity of the optimization problem and
generally will end up with problems such as multiple local optima and
potentially NP hardness. Recall that the nonnegative garrote estimate
of $\beta_i$ is $\widehatt{\beta}^\mathrm{LS}_i\theta_i(M)$. Since
$\widehatt{\beta}^\mathrm{LS}_i\neq0$ with probability one, $X_i$ will
be selected if and only if scaling factor $\theta_i>0$, in which case
$\theta_i$ behaves more or less like an indicator of the inclusion of
$X_i$ in the selected model. Therefore, the strong heredity principles
can be enforced by requiring
%
\begin{equation}
\label{strong}
\theta_{i}\le\theta_j \qquad \forall j\in\mathcaligr D_i.
\end{equation}
Note that if $\theta_i>0$, (\ref{strong}) will force the scaling
factor for all its parents to be positive and consequently active.
Since these constraints are linear in terms of the scaling factor,
minimizing~(\ref{shrink}) under~(\ref{strong}) remains a quadratic
program. Figure~\ref{fig:strong} illustrates the feasible region of
the nonnegative garrote with such constraints in contrast with the
original nonnegative garrote where no heredity rules are enforced. We
consider two effects and their interaction with the corresponding
shrinking factors denoted by $\theta_1$, $\theta_2$ and $\theta
_{12}$, respectively. In both situations the feasible region is a convex
polyhedron in the three dimensional space.

\begin{figure}[b]

\includegraphics{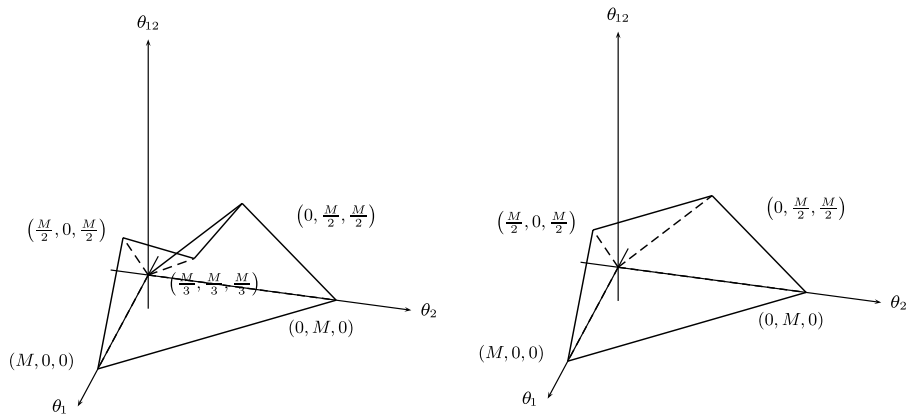}

\caption{Feasible region of the nonnegative garrote with constraint
$\theta_{12}\le\max(\theta_1,\theta_2)$ (left) and the relaxed
constraint $\theta_{12}\le\theta_1+\theta_2$ (right).}
\label{fig:weak}
\end{figure}

\subsection{Weak heredity principles}
Similarly, when considering weak heredity principles, we can require that
%
\begin{equation}
\label{weaknon}
\theta_i\le\max_{j\in\mathcaligr D_i} \theta_j.
\end{equation}
However, the feasible region under such constraints is no longer convex
as demonstrated in the left panel of Figure~\ref{fig:weak}.
Subsequently, minimizing~(\ref{shrink}) subject to~(\ref{weaknon}) is
not feasible. To overcome this problem, we suggest using the convex
envelop of these constraints for the weak heredity principles:
%
\begin{equation}
\label{weak}
\theta_{i}\le\sum_{j\in\mathcaligr D_i}\theta_j.
\end{equation}
Again, these constraints are linear in terms of the scaling factor and
minimizing (\ref{shrink}) under (\ref{weak}) remains a quadratic
program. Note that $\theta_i>0$ implies that $\sum_{j\in\mathcaligr D
_i}\theta_j>0$ and, therefore, (\ref{weak}) will require at least one
of its parents to be included in the model. In other words, constraint
(\ref{weak}) can be employed in place of (\ref{weaknon}) to enforce
the weak heredity principle. The small difference between the feasible
regions of (\ref{weaknon}) and (\ref{weak}) also suggests that the
selected model may only differ slightly between the two constraints. We
opt for (\ref{weak}) because of the great computational advantage it
brings about.

\subsection{Asymptotic properties}

To gain further insight to the proposed structured variable selection
and estimation methods, we study their asymptotic properties. We show
here that the proposed methods estimate the zero coefficients by zero
with probability tending to one, and at the same time give root-$n$
consistent estimate to the nonzero coefficients provided that the true
data generating mechanism satisfies such heredity principles. Denote by
$\mathcaligr A$ the indices of the predictors in the true model, that is,
$\mathcaligr A=\{j\dvtx \beta_j\neq0\}$. Write $\widehatt{\beta}^\mathrm{SVS}$ as
the estimate obtained from the proposed structured variable selection procedure.

Under strong heredity, the shrinkage factors $\hat\theta^{\mathrm{SVS}}(M)$ can be equivalently written in the Lagrange form [Boyd and
Vandenberghe (\citeyear{Boyd2004})]
\[
\arg\min_{\theta}\Biggl (\Vert Y-Z\theta\Vert^2+\lambda_n \sum
^p_{j=1}\theta_j \Biggr),
\]
subject to $\theta_j \ge0, \theta_j \le\min_{k\in\mathcaligr
D_j}\theta
_k$ for some Lagrange parameter $\lambda_n\ge0$. For the weak
heredity principle, we replace the constraints $\theta_j \le\min
_{k\in\mathcaligr D_j}\theta_k$ with $\theta_j \le\sum_{k\in
\mathcaligr D
_j}\theta_k$.

\begin{Theo}\label{thm1}
Assume that $X'X/n\to\Sigma$ and $\Sigma$ is positive definite. If
the true model satisfies the strong/weak heredity principles, and
$\lambda_n\to\infty$ in a fashion such that $\lambda_n=o(\sqrt
{n})$ as $n$ goes to $+\infty$, then the structured estimate with the
corresponding heredity principle satisfies $P(\widehatt{\beta}^\mathrm{SVS}_j=0)\to1$ for any $j\notin\mathcaligr A$, and $\widehatt{\beta
}^\mathrm{SVS}_j-\beta_j=O_p(n^{-1/2})$ if $j\in\mathcaligr A$.
\end{Theo}

All the proofs can be accessed as the supplement materials. Note that
when $\lambda_n=0$, there is no penalty and the proposed estimates
reduce to the least squares estimate which is consistent in estimation.
The theorems suggest that if instead the tuning parameter $\lambda_n$
escapes to infinity at a rate slower than $\sqrt{n}$, the resulting
estimates not only achieve root-$n$ consistency in terms of estimation
but also are consistent in variable selection, whereas the ordinary
least squares estimator does not possess such model selection ability.

\section{Generalized regression}\label{sec3}

The nonnegative garrote was originally introduced for variable
selection in multiple linear regression. But the idea can be extended
to more general regression settings where $Y$ depends on $X$ through a
scalar parameter $\eta(X)=\beta_0+X\beta,$ where $(\beta_0,\beta
')'$ is a $(p+1)$-dimensional unknown coefficient vector. It is worth
pointing out that such extensions have not been proposed in literature
so far.

A common approach to estimating $\eta$ is by means of the maximum
likelihood. Let $\ell(Y,\eta(X))$ be a negative log conditional
likelihood function of $Y|X$. The maximum likelihood estimate is given
as the minimizer of
\[
L(Y,\eta(X))=\sum_{i=1}^n \ell(y_i,\eta(\mathbf{x}_i)).
\]
For example, in logistic regression,
\[
\ell(y_i,\eta(\mathbf{x}_i))=y_i\eta(\mathbf{x}_i)-\log\bigl(1+e^{\eta
(\mathbf{x}
_i)} \bigr).
\]
More generally, $\ell$ can be replaced with any loss functions such
that its expectation $E(\ell(Y,\eta(X)))$ with respect to the joint
distribution of $(X,Y)$ is minimized at $\eta(\cdot)$.

To perform variable selection, we propose the following extension of
the original nonnegative garrote. We use the maximum likelihood
estimate $\widehatt{\beta}^\mathrm{MLE}$ as a preliminary estimate of
$\beta$. Similar to the original nonnegative garrote, define
$Z_j=X_j\widehatt{\beta}^\mathrm{MLE}_j$. Next we estimate the shrinkage
factors by
%
\begin{equation}
\label{gnng}
(\hat{\theta}_0,\widehatt{\theta}^\mathrm{SVS})'=\arg\min_{\theta
_0,\theta} L(Y,Z\theta+\theta_0),
\end{equation}
subject to $\sum\theta_j\le M$ and $\theta_j\ge0$ for any
$j=1,\ldots,p$. In the case of normal linear regression, $L$ becomes
the least squares and it is not hard to see that the solution of~(\ref
{gnng}) always satisfies $\widehatt{\theta}_0=0$ because all variables
are centered. Therefore, without loss of generality, we could assume
that there is no intercept in the normal linear regression. The same,
however, is not true for more general $L$ and, therefore, $\theta_0$
is included in (\ref{gnng}). Our final estimate of $\beta_j$ is then
given as $\widehatt{\beta}^\mathrm{MLE}_j\widehatt{\theta}_j^\mathrm{SVS}(M)$
for $j=1,\ldots,p$. To impose the strong or weak heredity
principle, we add additional constraints $\theta_j\le\min_{k\in
\mathcaligr D_j}\theta_k$ or $\theta_j\le\sum_{k\in\mathcaligr
D_j}\theta_k$,
respectively.

Theorem~\ref{thm1} can also be extended to more general regression settings.
Similar to before, under strong heredity,
\[
\widehatt{\theta}^\mathrm{SVS}(M)=\arg\min_{\theta_0,\theta}
\Biggl(L(Y,Z\theta+\theta_0)+\lambda_n \sum^p_{j=1}\theta_j \Biggr) ,
\]
subject to $\theta_j \ge0, \theta_j \le\min_{k\in\mathcaligr
D_j}\theta
_k$ for some $\lambda_n\ge0$. Under weak heredity principles, we use
the constraints $\theta_j \le\sum_{k\in\mathcaligr D_j}\theta_k$ instead
of $\theta_j \le\min_{k\in\mathcaligr D_j}\theta_k$.

We shall assume that the following regularity conditions hold:
\begin{longlist}
\item[(A.1)] $\ell(\cdot,\cdot)$ is a strictly convex function of
the second argument;
\item[(A.2)] the maximum likelihood estimate $\widehatt{\beta}^\mathrm{MLE}$ is root-$n$ consistent;
\item[(A.3)] the observed information matrix converges to a positive
definite matrix, that is,
%
\begin{equation}\label{C2}
\frac{1}{n}\sum_{i=1}^n \mathbf{x}_i \mathbf{x}_i'\ell
''(y_i,\mathbf{x}_i \widehatt{\beta
}^\mathrm{MLE}+\widehatt{\beta}^\mathrm{MLE}_0) \rightarrow_p {\Sigma},
\end{equation}
where ${\Sigma}$ is a positive definite matrix.
\end{longlist}

\begin{Theo}\label{thm2}
Under regularity conditions \textup{(A.1)--(A.3)}, if $\lambda_n\to\infty$ in
a fashion such that $\lambda_n=o(\sqrt{n})$ as $n$ goes to $+\infty
$, then $P(\widehatt{\beta}^\mathrm{SVS}_j=0)\to1$ for any\vspace*{-1pt} $j\notin
\mathcaligr A$, and $\widehatt{\beta}^\mathrm{SVS}_j-\beta
_j=O_p(n^{-1/2})$ if
$j\in\mathcaligr A$ provided that the true model satisfies\vspace*{1pt} the same heredity
principles.
\end{Theo}

\section{Computation}\label{sec4}

Similar to the original nonnegative garrote, the proposed structured
variable selection and estimation procedure proceeds in two steps.
First the solution path indexed by the tuning parameter $M$ is
constructed. The second step, oftentimes referred to as tuning, selects
the final estimate on the solution path.

\subsection{Linear regression}
We begin with linear regression. For both types of heredity principles,
the shrinkage factors for a given $M$ can be obtained from solving a
quadratic program of the following form:
%
\begin{equation}
\label{qpsvs}
\min_\theta\biggl(\frac{1}{2}\Vert Y-Z\theta\Vert ^{2} \biggr)\qquad \mbox{subject to } \sum
_{j=1}^p \theta_j\le M \mbox{ and }H\theta
\succeq{\mathbf0},
\end{equation}
where $H$ is a $m\times p$ matrix determined by the type of heredity
principles, ${\mathbf0}$ is a vector of zeros, and $\succeq$ means
``greater than or equal to'' in an element-wise manner. Equation~(\ref{qpsvs})
can be solved efficiently using standard quadratic programming
techniques, and the solution path of the proposed structured variable
selection and estimation procedure can be approximated by solving (\ref
{qpsvs}) for a fine grid of $M$'s.

Recently, Yuan and Lin (\citeyear{Yuan2006,YuanLin2007}) showed that the solution path of
the original nonnegative garrote is piecewise linear, and used this to
construct an efficient algorithm for building its whole solution path.
The original nonnegative garrote corresponds to the situation where the
matrix $H$ of (\ref{qpsvs}) is a $p\times p$ identity matrix. Similar
results can be expected for more general scenarios including the
proposed procedures, but the algorithm will become considerably more
complicated and running quadratic programming for a grid of tuning
parameter tends to be a computationally more efficient alternative.

Write $\hat{B}^\mathrm{LS}=\operatorname{diag}(\hat{\beta}_1^\mathrm{LS},\ldots
,\hat{\beta}_p^\mathrm{LS})$. The objective function of (\ref{qpsvs})
can be expressed as
\[
\tfrac{1}{2}\Vert Y-Z\theta\Vert ^{2} =\tfrac{1}{2} (Y'Y - 2 \theta'\hat
{B} X'Y + \theta'\hat{B} X'X \hat{B}\theta).
\]
Because $Y'Y$ does not depend on $\theta'$s, (\ref{qpsvs}) is
equivalent to
%
\begin{eqnarray}
&&\min_\theta\biggl(- \theta'\hat{B} X'Y + \frac{1}{2}\theta'\hat{B}
X'X \hat{B}\theta\biggr)\nonumber\\[-8pt]\\[-8pt]
&&\qquad \mbox{subject to } \sum
_{j=1}^p \theta_j\le M \mbox{ and }H\theta\succeq{\mathbf0},\nonumber
\end{eqnarray}
which depends on the sample size $n$ only through $X'Y$ and the Gram
matrix $X'X$. Both quantities are already computed in evaluating the
least squares. Therefore, when compared with the ordinary least squares
estimator, the additional computational cost of the proposed estimating
procedures is free of sample size $n$.

Once the solution path is constructed, our final estimate is chosen on
the \mbox{solution} path according to certain criterion. Such criterion often
reflects the prediction accuracy, which depends on the unknown
parameters and needs to be estimated. A commonly used criterion is the
multifold cross validation (CV). Multifold CV can be used to estimate
the prediction error of an estimator. The data $\mathcaligr L=\{
(y_i,\mathbf{x}
_i)\dvtx i=1,\ldots, n\}$ are first equally split into\vspace*{1pt} $V$ subsets
$\mathcaligr L_1,\ldots,\mathcaligr L_V$. Using the proposed method, and
data $L^{(v)}=\mathcaligr L-\mathcaligr L_v$, construct estimate
$\widehatt
{\beta}^{(v)}(M)$. The CV estimate of the prediction error is
\[
\wPE(\widehatt{\beta}(M))=\sum_v\sum_{(y_i,\mathbf
{x}_i)\in
\mathcaligr L_v} \bigl(y_i-\mathbf{x}_i\widehatt{\beta}^{(v)}(M) \bigr)^2.
\]
We select the tuning parameter $M$ by minimizing $\wPE(\widehatt{\beta}(M))$.
It is often suggested to use $V=10$ in
practice [Breiman (\citeyear{Breiman1995})].

It is not hard to see that $\wPE(\widehatt{\beta}^{(v)}(M))$ estimates
\[
\wPE(\widehatt{\beta}(M))=n\sigma^2+n \bigl(\beta-\widehatt{\beta
}(M) \bigr)'E(X'X) \bigl(\beta-\widehatt{\beta}(M) \bigr).
\]
Since the first term is the inherent prediction error due to the noise,
one often measures the goodness of an estimator using only the second
term, referred to as the model error:
%
\begin{equation}
\label{me}
\operatorname{ME}(\widehatt{\beta}(M))= \bigl(\beta-\widehatt{\beta}(M)
\bigr)'E(X'X) \bigl(\beta-\widehatt{\beta}(M) \bigr).
\end{equation}
Clearly, we can estimate the model error as $\wPE(\widehatt
{\beta}(M))/n-\widehat{\sigma^2}$, where $\widehat{\sigma^2}$ is
the noise variance estimate obtained from the ordinary least squares
estimate using all predictors.

\subsection{Generalized regression}
Similarly for more general regression settings, we solve
%
\begin{equation}
\label{gqpsvs}
\min_{\theta_0,\theta} L(Y,Z\theta+\theta_0)\qquad \mbox{subject to }
\sum_{j=1}^p \theta_j\le M \mbox{ and }H\theta\succeq{\mathbf0}
\end{equation}
for some matrix $H$. This can be done in an iterative fashion provided
that the loss function $L$ is strictly convex in its second argument.
At each iteration, denote $(\theta^{[0]}_0,\theta^{[0]})$ the
estimate from the previous iteration. We now approximate the objective
function using a quadratic function around $(\theta^{[0]}_0,\theta
^{[0]})$ and update the estimate by minimizing
\begin{eqnarray*}
&&\sum_{i=1}^n \biggl(\ell'\bigl(y_i,\mathbf{z}_i\theta^{[0]}+\theta
^{[0]}_0\bigr) \bigl[\mathbf{z}_i\bigl(\theta-\theta^{[0]}\bigr)+\bigl(\theta_0-
\theta^{[0]}_0\bigr) \bigr] \\
&&\qquad{}+{\frac{1}{2}} \ell''\bigl(y_i,\mathbf{z}_i\theta^{[0]}+\theta
^{[0]}_0\bigr) \bigl[\mathbf{z}_i\bigl(\theta-\theta^{[0]}\bigr)+\bigl(\theta_0-
\theta^{[0]}_0\bigr) \bigr]^2 \biggr) ,
\end{eqnarray*}
subject to $\sum\theta_j\le M$ and $H\theta\succeq{\mathbf0}$, where
the derivatives are taken with respect to the second argument of $\ell
$. Now it becomes a quadratic program. We repeat this until a certain
convergence criterion is met.

In choosing the optimal tuning parameter $M$ for general regression, we
again use the multifold cross-validation. It proceeds in the same
fashion as before except that we use a loss-dependent cross-validation score:
\[
\sum_v\sum_{(y_i,\mathbf{x}_i)\in\mathcaligr L_v}\ell\bigl(y_i,\mathbf
{x}_i\widehatt
{\beta}^{(v)}(M)+\widehatt{\beta}^{(v)}_0(M)\bigr).
\]

\section{Simulations}\label{sec5}

In this section we investigate the finite sample properties of the
proposed estimators. To fix ideas, we focus our attention on the usual
normal linear regression.

\subsection{Effect of structural constraints}
We first consider a couple of models that represent different scenarios
that may affect the performance of the proposed methods. In each of the
following models, we consider three explanatory variables $X_1,X_2,X_3$
that follow a multivariate normal distribution with $\operatorname{cov}(X_i,X_j)=
\rho^{|i-j|}$ with three different values for $\rho$:
$0.5, 0$ and $-0.5$. A quadratic model with nine terms
\[
Y = \beta_1X_1+\beta_2X_2+\beta_3X_3+\beta_{11}X_1^2+\beta
_{12}X_1X_2 + \cdots+\beta_{33}X_3^2+\varepsilon
\]
is considered. Therefore, we have a total of nine predictors, including
three main effects, three quadratic terms and three two-way
interactions. To demonstrate the importance of accounting for potential
hierarchical structure among the predictor variables, we apply the
nonnegative garrote estimator that recognizes strong heredity, weak
heredity and without heredity constraints. In particular, we enforce
the strong heredity principle by imposing the following constraints:
\begin{eqnarray*}
\theta_{11}
&\le&
\theta_1,\qquad \theta_{12}\le\theta_1,\qquad\theta_{12}\le\theta_2,\qquad \theta_{13}\le\theta_1,\qquad
\theta_{13}
\le\theta_3,\\
\theta_{22}
&\le&
\theta_2,\qquad \theta_{23}\le\theta_2,\qquad
\theta_{23}\le\theta_3,\qquad \theta_{33}\le\theta_3.
\end{eqnarray*}
To enforce the weak heredity, we require that
\begin{eqnarray*}
\theta_{11}
&\le&
\theta_1, \qquad\theta_{12}\le\theta_1+\theta_2,\qquad \theta_{13}\le\theta_1+\theta_3,\\
\theta_{22}
&\le&
\theta_2,\qquad \theta_{23}\le\theta_2+\theta_3,\qquad \theta_{33}\le\theta_3.
\end{eqnarray*}

We consider two data-generating models, one follows the strong heredity
principles and the other follows the weak heredity principles:

\begin{longlist}
\item[Model I.] The first model follows the strong heredity principle:
%
\begin{equation}
\label{mod1}
Y=3X_1+2X_2+1.5X_1X_2+\varepsilon;
\end{equation}
\item[Model II.] The second model is similar to Model I except that
the true data generating mechanism now follows the weak heredity principle:
%
\begin{equation}
\label{mod2}
Y=3X_1+2X_1^2+1.5X_1X_2+\varepsilon .
\end{equation}
\end{longlist}
For both models, the regression noise $\varepsilon\sim N(0,3^2)$.

For each model, 50 independent observations of $(X_1,X_2,X_3,Y)$ are
collected, and a quadratic model with nine terms is analyzed. We choose
the tuning parameter by ten-fold cross-validation as described in the
last section. Following Breiman (\citeyear{Breiman1995}), we use the model error (\ref
{me}) as the gold standard in comparing different methods. We repeat
the experiment for 1000 times for each model and the results are
summarized in Table~\ref{tab:ex1}. The numbers in the parentheses are
the standard errors. We can see that the model errors are smaller for
both weak and strong heredity models compared to the model that does
not incorporate any of the heredity principles. Paired $t$-tests
confirmed that most of the observed reductions in model error are
significant at the 5\% level.

\begin{table}[b]
\tablewidth=9.2cm
\caption{Model error comparisons. Model \textup{I} satisfies strong heredity
and Model \textup{II} satisfies weak heredity}
\label{tab:ex1}
\begin{tabular*}{9.2cm}{@{\extracolsep{\fill}}lccc@{}}
\hline
& \textbf{No heredity} & \textbf{Weak heredity} & \textbf{Strong heredity} \\
\hline
\multicolumn{4}{c}{Model I}\\
$\rho=0.5$ & 1.79 & 1.70 & 1.59 \\
& (0.05) & (0.05) & (0.04) \\
$\rho=0$ & 1.57 & 1.56 & 1.43 \\
& (0.04) & (0.04) & (0.04) \\
$\rho=-0.5$ & 1.78 & 1.69 & 1.54 \\
& (0.05) & (0.04) & (0.04) \\[6pt]
\multicolumn{4}{c}{Model II}\\
$\rho=0.5$ & 1.77 & 1.61 & 1.72 \\
& (0.05) & (0.05) & (0.04) \\
$\rho=0$ & 1.79 & 1.53 & 1.70 \\
& (0.05) & (0.04) & (0.04) \\
$\rho=-0.5$ & 1.79 & 1.68 & 1.76 \\
& (0.04) & (0.04) & (0.04) \\
\hline
\end{tabular*}
\end{table}

For Model I, the nonnegative garrote that respects the strong heredity
principles enjoys the best performance, followed by the one with weak
heredity principles. This example demonstrates the benefit of
recognizing the effect heredity. Note that the model considered here
also follows the weak heredity principle, which explains why the
nonnegative garrote estimator with weak heredity outperforms the one
that does not enforce any heredity constraints. For Model II, the
nonnegative garrote with weak heredity performs the best.
Interestingly, the nonnegative garrote with strong heredity performs
better than the original nonnegative garrote. One possible explanation
is that the reduced feasible region with strong heredity, although
introducing bias, at the same time makes tuning easier.

To gain further insight, we look into the model selection ability of
the structured variable selection. To separate the strength of a method
and effect of tuning, for each of the simulated data, we check whether
or not there is any tuning parameter such that the corresponding
estimate conforms with the true model. The frequency for each method to
select the right model is given in Table~\ref{tab:ex1f}, which clearly
shows that the proposed structured variable selection methods pick the
right models more often than the original nonnegative garrote. Note
that the strong heredity version of the method can never pick Model II
correctly as it violates the strong heredity principle. We also want to
point out that such comparison, although useful, needs to be understood
with caution. In practice, no model is perfect and selecting an
additional main effect $X_2$ so that Model II can satisfy strong
heredity may be a much more preferable alternative to many.
%
\begin{table}
\caption{Frequency of selecting the right model}
\label{tab:ex1f}
\begin{tabular}{@{}lccc@{}}
\hline
& \textbf{No heredity} & \textbf{Weak heredity} & \textbf{Strong heredity} \\
\hline
\multicolumn{4}{c}{Model I}\\
$\rho=0.5$ & 65.5\% & 71.5\% & 82.0\% \\
$\rho=0$ & 85.0\% & 86.5\% & 90.5\% \\
$\rho=-0.5$ & 66.5\% & 73.5\% & 81.5\% \\[6pt]
\multicolumn{4}{c}{Model II}\\
$\rho=0.5$ & 65.5\% & 75.5\% & 0.00\% \\
$\rho=0$ & 83.0\% & 90.0\% & 0.00\% \\
$\rho=-0.5$ & 56.5\% & 72.5\% & 0.00\% \\
\hline
\end{tabular}
\end{table}

\begin{figure}

\includegraphics{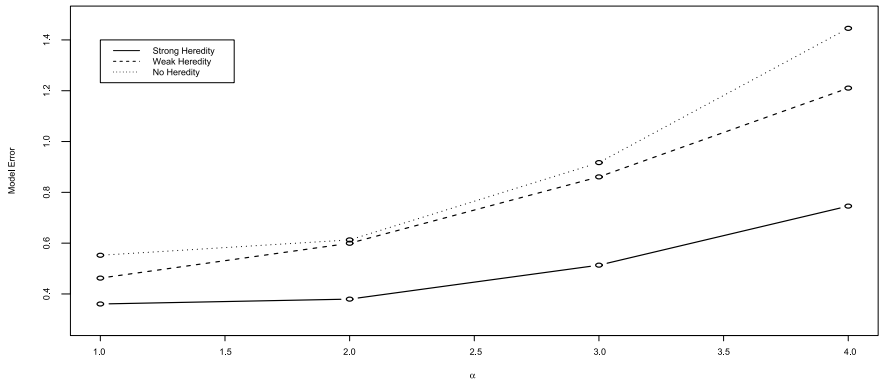}

\caption{Effect of the magnitude of the interactions.}
\label{fig:revsim1}
\end{figure}

We also checked how effective the ten-fold cross-validation is in
picking the right model when it does not follow any of the heredity
principles. We generated the data from the model
\[
Y=3X_1+2X_1^2+1.5X_2^2+\varepsilon,
\]
where the set up for simulation remains the same as before. Note that
this model does not follow any of the heredity principles. For each
run, we ran the nonnegative garrote with weak heredity, strong heredity
and no heredity. We chose the best among these three estimators using
ten-fold cross-validation. Note that the three estimators may take
different values of the tuning parameter. Among 1000 runs, 64.1\% of
the time, nonnegative garrote with no heredity principle was elected.
In contrast, for either Model I or Model II with a similar setup, less
than 10\% of the time nonnegative garrote with no heredity principle
was elected. This is quite a satisfactory performance.

\subsection{Effect of the size of the interactions}
The next example is designed to illustrate the effect of the magnitude
of the interaction on the proposed methods. We use a similar setup as
before but now with four main effects $X_1,X_2,X_3, X_4$, four
quadratic terms and six two-way interactions. The true data generating
mechanism is given by
%
\begin{equation}
\label{eq:effectsize}
Y=3X_1+2X_2+1.5X_3+\alpha(X_1X_2-X_1X_3)+\varepsilon,
\end{equation}
where $\alpha=1,2,3,4$ and $\varepsilon\sim N(0,\sigma^2)$ with
$\sigma
^2$ chosen so that the signal-to-noise ratio is always $3\dvtx 1$. Similar to
before, the sample size $n=50$. Figure~\ref{fig:revsim1} shows the
mean model error estimated over 1000 runs. We can see that the strong
and weak heredity models perform better than the no heredity model and
the improvement becomes more significant as the strength of the
interaction effect increases.

\subsection{Large $p$}
\begin{figure}

\includegraphics{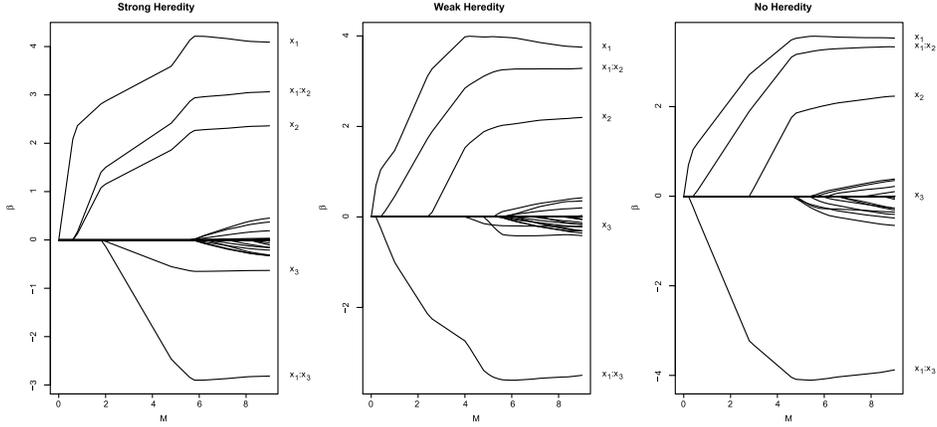}

\caption{Simulation when $p>n$: solution for different versions of the
nonnegative garrote.}
\label{fig:aoasrevsim}
\end{figure}

To fix the idea, we have focused on using the least squares estimator
as our initial estimator. The least squares estimators are known to
perform poorly when the number of predictors is large when compared
with the sample size. In particular, it is not applicable when the
number of predictors exceeds the sample size. However, as shown in Yuan
and Lin (\citeyear{YuanLin2007}), other initial estimators can also be used. In
particular, they suggested ridge regression as one of the alternatives
to the least squares estimator. To demonstrate such an extension, we
consider again the regression model (\ref{eq:effectsize}) but with ten
main effects $X_1,\ldots, X_{10}$ and ten quadratic terms, as well as
45 interactions. The total number of effects ($p=65$) exceeds the
number of observations ($n=50$) and, therefore, the ridge regression
tuned with GCV was used as the initial estimator. Figure \ref
{fig:aoasrevsim} shows the solution path of the nonnegative garrote
with strong heredity, weak heredity and without any heredity for a
typical simulated data with $\alpha=4$.

It is interesting to notice from Figure \ref{fig:aoasrevsim} that the
appropriate heredity principle, in this case strong heredity, is
extremely valuable in distinguishing the true effect~$X_3$ from other
spurious effects. This further confirms the importance of heredity principles.

\section{Real data examples}\label{sec6}

In this section we apply the methods from Section~\ref{sec2} to several real
data examples.

\subsection{Linear regression example}
\begin{figure}

\includegraphics{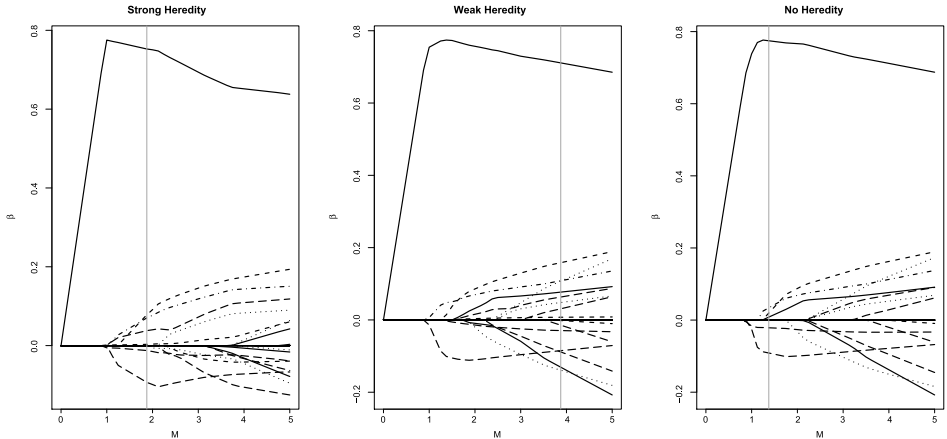}

\caption{Solution path for different versions of the nonnegative garrote.}
\label{fig:prostate}
\end{figure}

The first is the prostate data, previously used in Tibshirani (\citeyear{Tibshirani1996}).
The data consist of the medical records of $97$ male patients who were
about to receive a radical prostatectomy. The response \mbox{variable} is the
level of prostate specific antigen, and there are 8 explanatory
variables. The explanatory variables are eight clinical measures:
log(cancer volume) (lcavol), log(prostate weight) (lweight), age,
log(benign prostatic hyperplasia amount) (lbph), seminal vesicle
invasion (svi), log(capsular penetration) (lcp), Gleason score
(gleason) and percentage Gleason scores 4 or 5 (pgg45). We consider
model~(\ref{2way}) with main effects, quadratic terms and two way
interactions, which gives us a total of 44 predictors. Figure~\ref{fig:prostate}
gives the solution path of the nonnegative garrote with
strong heredity, weak heredity and without any heredity constraints.
The vertical grey lines represent the models that are selected by the
ten-fold cross-validation.

\begin{figure}[b]

\includegraphics{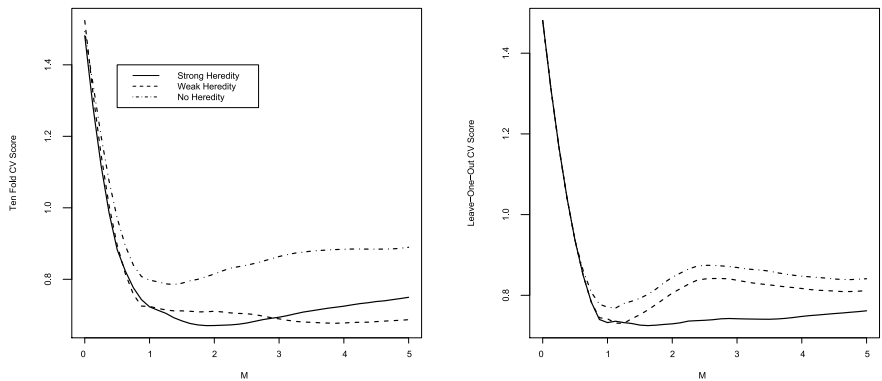}

\caption{Cross-validation scores for the prostate data.}
\label{fig:prostate-cv}
\end{figure}

To determine which type of heredity principle to use for the analysis,
we calculated the ten-fold cross-validation scores for each method. The
cross-validation scores as functions of the tuning parameter $M$ are
given in the right panel of Figure~\ref{fig:prostate-cv}.

Cross-validation suggests the validity of heredity principles. The
strong heredity is particularly favored with the smallest score. Using
ten-fold cross-validation, the original nonnegative garrote that
neglects the effect heredity chooses a six variable model: \textit{lcavol,
lweight, lbph, gleason$^2$, lbph:svi} and \textit{svi:pgg45}. Note that
this model does not satisfy heredity principle, because \textit
{gleason$^2$} and \textit{svi:pgg45} are included without any of its parent
factors. In contrast, the nonnegative garrote with strong heredity
selects a model with seven variables: \textit{lcavol, lweight, lbph, svi,
gleason, gleason$^2$} and \textit{lbph:svi}. The model selected by the
weak heredity, although comparable in terms of cross validation score,
is considerably bigger with 16 variables. The estimated model errors
for the strong heredity, weak heredity and no heredity nonnegative
garrote are $0.18$, $0.19$ and $0.30$, respectively, which clearly
favors the methods that account for the effect heredity.

To further assess the variability of the ten-fold cross-validation, we
also ran the leave-one-out cross-validation on the data. The
leave-one-out scores are given in the right panel of Figure~\ref{fig:prostate-cv}.
It shows a similar pattern as the ten-fold
cross-validation. In what follows, we shall continue to use the ten-fold cross-validation because of the tremendous computational advantage
it brings about.

\subsection{Logisitic regression example}
\begin{figure}

\includegraphics{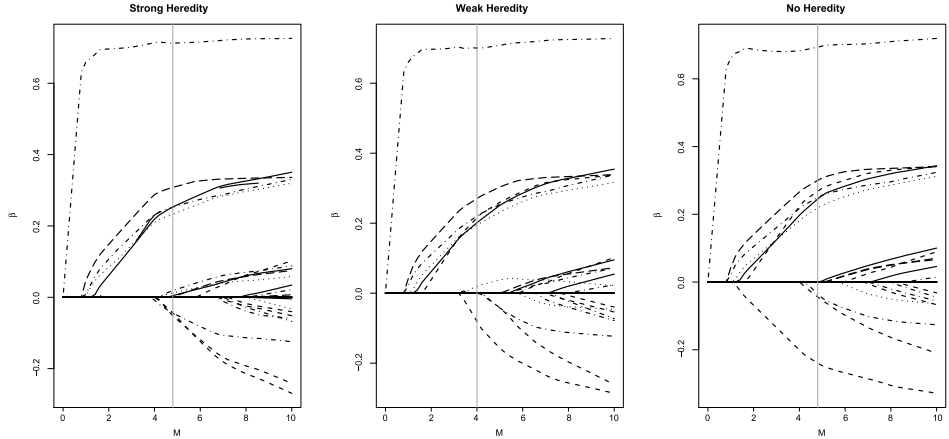}

\caption{Solution paths for the heart data.}
\label{fig:heart-path}
\end{figure}

To illustrate the strategy in more general regression settings, we
consider a logistic regression for the South African heart disease data
previously used in Hastie, Tibshirani and Friedman (\citeyear{Hastie2003}). The data
consist of 9 different measures of 462 subjects and the responses
indicating the presence of heart disease. We again consider a quadratic
model. There is one binary predictor which leaves a total of 53 terms.
Nonnegative garrote with strong heredity, weak heredity and without
heredity were applied to the data set. The solution paths are given in
Figure~\ref{fig:heart-path}.

\begin{figure}

\includegraphics{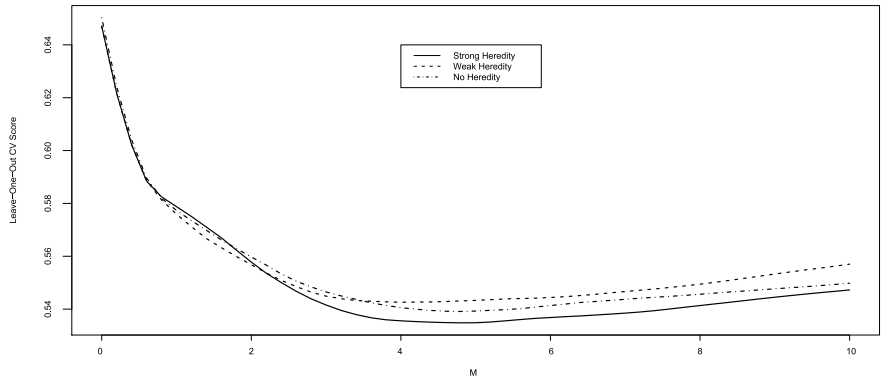}

\caption{Cross-validation scores for the heart data.}
\label{fig:heart-cv}
\end{figure}

The cross-validation scores for the three different methods are given
in Figure~\ref{fig:heart-cv}. As we can see from the figure,
nonnegative garrote with strong heredity principles achieves the lowest
cross-validation score, followed by the one without heredity principles.

\subsection{Prediction performance on several benchmark data}

To gain further insight on the merits of the proposed structured
variable selection and estimation techniques, we apply them to seven
benchmark data sets, including the previous two examples. The Ozone
data, originally used in Breiman and Friedman (\citeyear{Breiman1985}), consist of the
daily maximum one-hour-average ozone reading and eight meteorological
variables in the
Los Angeles basin for 330 days in 1976. The goal is to predict the
daily maximum one-hour-average ozone reading using the other eight
variables. The Boston housing data include statistics for 506 census
tracts of Boston from the 1970 census [Harrison and Rubinfeld (\citeyear{Harrison1978})].
The problem is to predict the median value of owner-occupied homes
based on 13 demographic and geological measures. The Diabetes data,
previously analyzed by Efron et al. (\citeyear{Efron2004}), consist of eleven clinical
measurements for a total of 442 diabetes patients. The goal is to
predict a quantitative measure of disease progression one year after
the baseline using the other ten measures that were collected at the
baseline. Along with the prostate data, these data sets are used to
demonstrate our methods in the usual normal linear regression setting.

To illustrate the performance of the structured variable selection and
estimation in more general regression settings, we include two other
logistic regression examples along with the South African Heart data.
The Pima Indians Diabetes data have 392 observations on nine variables.
The purpose is to predict whether or not a particular subject has
diabetes using eight remaining variables. The BUPA Liver Disorder data
include eight variables and the goal is to relate a binary response
with seven clinical measurements. Both data sets are available from the
UCI Repository of machine learning databases [Newman et al. (\citeyear{Newman1998})].

We consider methods that do not incorporate heredity principles or
respect weak or strong heredity principles. For each method, we
estimate the prediction error using ten-fold cross-validation, that is,
the mean squared error in the case of the four linear regression
examples, and the misclassification rate in the case of the three
classification examples. Table \ref{tab:real} documents the findings.
Similar to the Heart data we discussed earlier, the total number of
effects ($p$) can be different for the same number of main effects
($q$) due to the existence of binary variables. As the results from
Table~\ref{tab:real} suggest, incorporating the heredity principles
leads to improved prediction for all seven data sets. Note that for the
four regression data sets, the prediction error depends on the scale of
the response and therefore should not be compared across data sets. For
example, the response variable of the diabetes data ranges from 25 to
346 with a variance of 5943.331. In contrast, the response variable of
the prostate data ranges from $-0.43$ to 5.48 with a variance of 1.46.

\begin{table}
\caption{Prediction performance on seven real data sets}
\label{tab:real}
\begin{tabular*}{\textwidth}{@{\extracolsep{\fill}}ld{3.0}d{2.0}d{3.0}d{4.3}cc@{}}
\hline
\textbf{Data} & \multicolumn{1}{c}{{\textbf{\textit n}}}
& \multicolumn{1}{c}{{\textbf{\textit q}}}
& \multicolumn{1}{c}{{\textbf{\textit p}}}
& \multicolumn{1}{c}{\textbf{No heredity}} & \multicolumn{1}{c}{\textbf{Weak heredity}}
& \multicolumn{1}{c@{}}{\textbf{Strong heredity}}\\
\hline
Boston & 506 & 13 & 103 & 12.609 & \multicolumn{1}{c}{\phantom{\textbf{87}}\textbf{12.403}} & \phantom{16}12.661\\
Diabetes & 442 & 10 & 64 & 3077.471 & \multicolumn{1}{c}{\textbf{2987.447}} & 3116.989\\
Ozone & 203 & 9 & 54 & 16.558 & \multicolumn{1}{c}{\phantom{\textbf{87}}\textbf{15.100}} & \phantom{16}15.397\\
Prostate & 97 & 8 & 44 & 0.624 & \phantom{\textbf{987}}0.632 & \phantom{116}\textbf{0.584}\\[3pt]
BUPA & 345 & 6 & 27 & 0.287 & \phantom{\textbf{987}}0.279 & \phantom{116}\textbf{0.267}\\
Heart & 462 & 9 & 53 & 0.286 & \phantom{\textbf{987}}0.275 & \phantom{116}\textbf{0.262}\\
Pima & 392 & 8 & 44 & 0.199 & \phantom{\textbf{987}}0.214 & \phantom{116}\textbf{0.196}\\
\hline
\end{tabular*}
\end{table}

\section{Discussions}\label{sec7}

When a large number of variables are entertained, variable selection
becomes important. With a number of competing models that are virtually
indistinguishable in fitting the data, it is often advocated to select
a model with the smaller number of variables. But this principle alone
may lead to models that are not interpretable. In this paper we
proposed structured variable selection and estimation methods that can
effectively incorporate the hierarchical structure among the predictors
in variable selection and regression coefficient estimation. The
proposed methods select models that satisfy the heredity principle and
are much more interpretable in practice. The proposed methods adopt the
idea of the nonnegative garrote and inherit its advantages. They are
easy to compute and enjoy good theoretical properties.

Similar to the original nonnegative garrote, the proposed method
involves the choice of a tuning parameter which also amounts to the
selection of a final model. Throughout the paper, we have focused on
using the cross-validation for such a purpose. Other tuning methods
could also be used. In particular, it is known that prediction-based
tuning may result in unnecessarily large models. Several heuristic
methods are often adopted in practice to alleviate such problems. One
of the most popular choices is the so-called one standard error rule
[Breiman et al. (\citeyear{Breiman1984})], where instead of choosing
the model that minimizes the cross-validation score, one chooses the
simplest model with a cross-validation score within one standard error
from the smallest. Our experience also suggests that a visual
examination of the solution path and the cross-validation scores often
leads to further insights.

The proposed method can also be used in other statistical problems
whenever the structures among predictors should be respected in model
building. In some applications, certain predictor variables may be
known apriori to be more important than the others. This may happen,
for example, in time series prediction where more recent observations
generally should be more predictive of future observations.

\printaddresses


\begin{thebibliography}{99}

%
\bibitem[\protect\citeauthoryear{}{2004}]{Boyd2004}
\textsc{Boyd, S.} and \textsc{Vandenberghe, L.} (2004). \textit{Convex
Optimization}.
Cambridge Univ. Press, Cambridge.
\MR{2061575}
%
\bibitem[\protect\citeauthoryear{}{1995}]{Breiman1995}
\textsc{Breiman, L.} (1995). Better subset regression using the
nonnegative garrote. \textit{Technometrics} \textbf{37} 373--384.
\MR{1365720}
%
\bibitem[\protect\citeauthoryear{}{1985}]{Breiman1985}
\textsc{Breiman, L.} and \textsc{Friedman, J.} (1985).
Estimating optimal
transformations for multiple regression and correlation. \textit{J. Amer. Statist. Assoc.}
\textbf{80} 580--598.
\MR{0803258}
%
\bibitem[\protect\citeauthoryear{}{1984}]{Breiman1984}
\textsc{Breiman, L., Friedman, J., Stone, C.} and \textsc
{Olshen, R.} (1984). \textit{Classifcation and Regression Trees}. Chapman \& Hall/CRC, New
York.
\MR{0726392}
%
\bibitem[\protect\citeauthoryear{}{1996}]{Chipman1996}
\textsc{Chipman, H.} (1996). Bayesian variable selection with related
predictors. \textit{Canad. J. Statist.} \textbf{24} 17--36.
\MR{1394738}
%
\bibitem[\protect\citeauthoryear{}{1997}]{Chipman1997}
\textsc{Chipman, H., Hamada, M.} and \textsc{Wu, C. F. J.}
(1997). A Bayesian variable
selection approach for analyzing designed experiments with complex
aliasing. \textit{Technometrics} \textbf{39} 372--381.
%
\bibitem[\protect\citeauthoryear{}{2006}]{Choi2006}
\textsc{Choi, N., Li, W.} and \textsc{Zhu, J.} (2006). Variable
selection with the
strong heredity constraint and its oracle property. Technical report.
%
\bibitem[\protect\citeauthoryear{}{2004}]{Efron2004}
\textsc{Efron, B., Johnstone, I., Hastie, T.} and \textsc
{Tibshirani, R.} (2004).
Least angle regression (with discussion). \textit{Ann.
Statist.} \textbf{32} 407--499.
\MR{2060166}
%
\bibitem[\protect\citeauthoryear{}{2001}]{Fan2001}
\textsc{Fan, J.} and \textsc{Li, R.} (2001). Variable selection
via nonconcave
penalized likelihood and its oracle properties. \textit{J.
Amer. Statist. Assoc.} \textbf{96} 1348--1360.
\MR{1946581}
%
\bibitem[\protect\citeauthoryear{}{1993}]{George1993}
\textsc{George, E. I.} and \textsc{McCulloch, R. E.} (1993).
Variable selection via
Gibbs sampling. \textit{J. Amer. Statist. Assoc.}
\textbf{88} 881--889.
%
\bibitem[\protect\citeauthoryear{}{1992}]{Hamada1992}
\textsc{Hamada, M.} and \textsc{Wu, C. F. J.} (1992).
Analysis of designed
experiments with complex aliasing. \textit{Journal of Quality Technology}
\textbf{24} 130--137.
%
\bibitem[\protect\citeauthoryear{}{1978}]{Harrison1978}
\textsc{Harrison, D.} and \textsc{Rubinfeld, D.} (1978).
Hedonic prices and the
demand for clean air. \textit{Journal of Environmental Economics and
Management} \textbf{5} 81--102.
%
\bibitem[\protect\citeauthoryear{}{2003}]{Hastie2003}
\textsc{Hastie, T., Tibshirani, R.} and \textsc{Friedman, J.}
(2003). \textit{The
Elements of Statistical Learning: Data Mining, Inference, and
Prediction}. Springer, New York.
\MR{1851606}
%
\bibitem[\protect\citeauthoryear{}{2007}]{Joseph2007}
\textsc{Joseph, V. R.} and \textsc{Delaney, J. D.} (2007).
Functionally induced
priors for the analysis of experiments. \textit{Technometrics}
\textbf{49} 1--11.
\MR{2345447}
%
\bibitem[\protect\citeauthoryear{}{2006}]{Li2006}
\textsc{Li, X., Sundarsanam, N.} and \textsc{Frey, D.} (2006).
Regularities in data
from factorial experiments. \textit{Complexity} \textbf{11} 32--45.
%
\bibitem[\protect\citeauthoryear{}{2002}]{McCullagh2002}
\textsc{McCullagh, P.} (2002). What is a statistical model (with
discussion). \textit{Ann. Statist.} \textbf{30} 1225--1310.
\MR{1936320}
%
\bibitem[\protect\citeauthoryear{}{1989}]{McCullagh1989}
\textsc{McCullagh, P.} and Nelder, J. (1989). \textit{Generalized Linear
Models}, 2nd ed. Chapman \& Hall, London.
\MR{0727836}
%
\bibitem[\protect\citeauthoryear{}{1977}]{Nelder1977}
\textsc{Nelder, J.} (1977). A reformulation of linear models.
\textit{J. Roy. Statist. Soc. Ser. A} \textbf{140} 48--77.
\MR{0458743}
%
\bibitem[\protect\citeauthoryear{}{1994}]{Nelder1994}
\textsc{Nelder, J.} (1994). The statistics of linear models.
\textit{Statist. Comput.} \textbf{4} 221--234.
%
\bibitem[\protect\citeauthoryear{}{1998}]{Nelder1998}
\textsc{Nelder, J.} (1998). The selection of terms in response-surface
models---how strong is the weak-heredity principle? \textit{Amer.
Statist.} \textbf{52} 315--318.
%
\bibitem[\protect\citeauthoryear{}{1998}]{Newman1998}
\textsc{Newman, D., Hettich, S., Blake, C.} and \textsc{Merz,
C.} (1998). UCI
repository of machine learning databases. Dept. Information and Computer
Science, Univ. California,
Irvine, CA.
Available at \url{http://www.ics.uci.edu/\textasciitilde mlearn/MLRepository.html}.
%
\bibitem[\protect\citeauthoryear{}{2000}]{Osborne2000}
\textsc{Osborne, M., Presnell, B.} and \textsc{Turlach, B.}
(2000). A new approach
to variable selection in least squares problems. \textit{IMA J.
Numer. Anal.} \textbf{20} 389--403.
\MR{1773265}
%
\bibitem[\protect\citeauthoryear{}{1996}]{Tibshirani1996}
\textsc{Tibshirani, R.} (1996). Regression shrinkage and
selection via the
lasso. \textit{J. Roy. Statist. Soc. Ser. B}
\textbf{58} 267--288.
\MR{1379242}
%
\bibitem[\protect\citeauthoryear{}{2004}]{Turlach2004}
\textsc{Turlach, B.} (2004). Discussion of ``Least angle regression.''
\textit{Ann. Statist.} \textbf{32} 481--490.
\MR{2060166}
%
\bibitem[\protect\citeauthoryear{}{2007}]{Yuan2007}
\textsc{Yuan, M., Joseph, V. R.} and \textsc{Lin, Y.} (2007).
An efficient variable
selection approach for analyzing designed experiments. \textit
{Technometrics} \textbf{49} 430--439.
\MR{2414515}
%
\bibitem[\protect\citeauthoryear{}{2006}]{Yuan2006}
\textsc{Yuan, M.} and \textsc{Lin, Y.} (2006). Model selection
and estimation in
regression with grouped variables. \textit{J. Roy. Statist. Soc. Ser. B} \textbf{68} 49--67.
\MR{2212574}
%
\bibitem[\protect\citeauthoryear{}{2007}]{YuanLin2007}
\textsc{Yuan, M.} and \textsc{Lin, Y.} (2007). On the the
nonnegative garrote
estimator. \textit{J. Roy. Statist. Soc. Ser. B}
\textbf{69} 143--161.
\MR{2325269}

\bibitem[\protect\citeauthoryear{}{2009}]{Zhao2006}
\textsc{Zhao, P., Rocha, G.} and \textsc{Yu, B.} (2009).
The composite absolute penalties family for grouped and hierarchical
variable selection. \textit{Ann. Statist.} \textbf{37} 3468--3497.
\MR{2549566}

\end{thebibliography}
\end{document}